\documentclass[preprint]{aa}
\usepackage{txfonts}
\usepackage{graphicx,subfigure}
\usepackage{epsfig}
\usepackage{epsf}
\usepackage{lscape}
\usepackage{natbib}
\usepackage{rotating}
\usepackage{tabularx}
\usepackage{lscape}
\usepackage{supertabular}
\newcommand{\mic}{\,{\rm \mu m} }

\begin{document}
\title{Detection and characterization of a 500 $\mic$ dust emissivity excess in the Galactic plane using Herschel/Hi-GAL observations\thanks{{\it Herschel} is an ESA space observatory
  with science instruments provided by European-led Principal
  investigator consortia and with important participation from NASA.}}
%\subtitle{}

\author{D. Paradis \inst{1,2} \and  R. Paladini \inst{3}  \and
  A. Noriega-Crespo \inst{3} \and C. M\'eny \inst{1,2} \and F. Piacentini
  \inst{4} \and M. A. Thompson \inst{5} \and D. J. Marshall \inst{1,2} \and
M. Veneziani \inst{3} \and  J.-P. Bernard \inst{1,2} \and S. Molinari
\inst{6}} 
 
\institute{Universit\'e de Toulouse; UPS-OMP; IRAP; Toulouse, France 
\and
CNRS; IRAP; 9 Av. du Colonel Roche, BP 44346, F-31028, Toulouse, cedex
4, France
\and
Spitzer Science Center, California Institute of Technology, 1200
E. California Blvd, Pasadena, CA 91125, USA
\and
Dipartimento de Fisica, Universita di Roma 1 La Sapienza, Roma, Italy
\and
Centre for Astrophysics Research, Science and Technology Research
Institute, University of Hertfordshire, Hatfield, UK
\and
INAF-Instituto Fisica Spazio Interplanetario, Via Fosso del Cavaliere
100, I-00133 Roma, Italy}

\authorrunning{Paradis et al.}
\titlerunning{Detection and characterization of a $500\mic$ excess in the Galactic plane}

\abstract
{Past and recent observations have revealed unexpected variations
in the far-infrared - millimeter (FIR-mm) dust emissivity in the
interstellar medium. In the Herschel spectral range, those
are often referred to as a $500\mic$ emission excess. Several dust
emission models have been developed to interpret astrophysical data in the FIR-mm domain. 
However, these are commonly unable 
to fully reconcile theoretical predictions with observations. In contrast, the recently revised two level
  system (TLS) model, based on the
disordered internal structure of amorphous dust grains, seems to provide 
a promising way of interpreting existing data.} 
{The newly available Herschel infrared Galactic (Hi-GAL) data, which covers most of
the inner Milky Way, offers a unique opportunity to investigate possible        
variations in the dust emission properties both with wavelength and environment. The goal of our 
analysis is to constrain the internal structure of the largest dust
grains on Galactic scales, in the 
framework of the TLS model.}
{By combining the IRIS (Improved Reprocessing of the IRAS Survey) 100 $\mic$ with the Hi-GAL 160, 250, 350, and 500 $\mic$ data, we model the dust emission spectra in
  each pixel of the Hi-GAL maps, using both the TLS model and, for comparison, a single modified
black-body fit. The effect of temperature mixing along the line of
sight is investigated to test the robustness of our results.}  
{We find a slight decrease in the dust temperature with
distance from the Galactic center, confirming previous
results. We also report the detection of a significant 500 $\mic$
emissivity excess in the peripheral regions of the plane ($35\degr<|l|<70\degr$)
of about 13-15$\%$ of the emissivity, which can reach up to 20$\%$ in
some HII regions . We present the spatial distributions of the best-fit values for
the two main parameters of the TLS model, i.e. the charge correlation length,
$l_{c}$, used to characterize the disordered charge distribution
  (DCD) part of the model, and the amplitude $A$ of the TLS
processes with respect to the DCD effect. These distributions
illustrate the variations in the dust properties with environment, in
particular the plausible
existence of an overall gradient with distance to the Galactic
center. A comparison with previous findings in the solar neighborhood shows
that the local value of the excess is less than expected from
the Galactic gradient observed here.}
{}

\maketitle
\section{Introduction}
The interstellar medium (ISM) is where matter 
from diffuse clouds is brought into stars. In this context, studying dust evolution 
is important as variations in the properties of the ISM constituents likely affect 
the star formation process. In addition, understanding the emission
from big grains (BGs) is important in interpreting experiments devoted to the observation of the cosmic microwave background (CMB) and its fluctuations. 
These experiments typically operate in the millimeter domain, where the cosmological signal is dominated 
by the Galactic foreground emission, of which dust is one of the components.

The spectral energy distributions (SEDs) of BGs can be
approximated by modified black-body emission
\begin{equation}
I_{\nu}(\lambda) =\epsilon_{\nu} B_{\nu}(\lambda,T_d) N_H, 
\label{eq:Inu}
\end{equation}
where $\rm I_{\nu}(\lambda)$ is the brightness, $\epsilon_{\nu}$ is the dust
emissivity per hydrogen column, $\rm B_{\nu}$ is the
Planck function, T$\rm _d$ is the dust temperature, and $N_H$ is the
hydrogen column density. When Equation \ref{eq:Inu} is used to interpret
  the observed emission, $\epsilon_{\nu}$ is the mean dust
  emissivity along the line of sight (LOS). This approximation is justified because the
  medium is optically thin in the far-infrared (FIR) and submillimeter (submm). The dust temperature
  is however expected to vary along the LOS (see Section 4.2.2). The dust emissivity is usually defined as
\begin{equation}
\epsilon_{\nu}(\lambda)=\epsilon_{\nu}(\lambda_0) \left ( \frac{\lambda}{\lambda_0} \right )^{-\beta},
\end{equation}
where $\rm \epsilon_{\nu}(\lambda_0)$ is the emissivity at wavelength $\rm \lambda_0$, and $\rm \beta$ is
the emissivity spectral index, usually taken to be equal to 2. 
In recent years, however, an increasing number of studies have found that the emissivity might actually depart from a simple $\lambda^{-2}$ power-law. Indeed, the emissivity spectral index
seems to be temperature-dependent: its value has been shown to
decrease with increasing temperature \citep{Dupac03, Desert08,
  Veneziani10, Paradis10}. This $\rm T_d$-$\beta$ anti-correlation has
been extensively debated. Some authors claim that this behavior is only the consequence of mis-handling noisy data and/or temperature mixing along
the LOS \citep{Shetty09}. However, all the aforementioned studies have concluded that
the effect was not due to noise, and some analyses have included tests that
rejected temperature mixing as the cause of the observed trend \citep{Paradis09, Malinen11}.
We note that in some extreme cases, the inferred $\beta$ values can
reach up to 3.5-4 at low apparent temperatures ($\leq$10 K), which
appears very difficult to reconcile with a single value of $\beta$ and
temperature mixing effects. It appears more likely that actual variations
in the dust emissivity are at play.
\begin{figure*}[!t]
\begin{center}
\includegraphics[width=14cm]{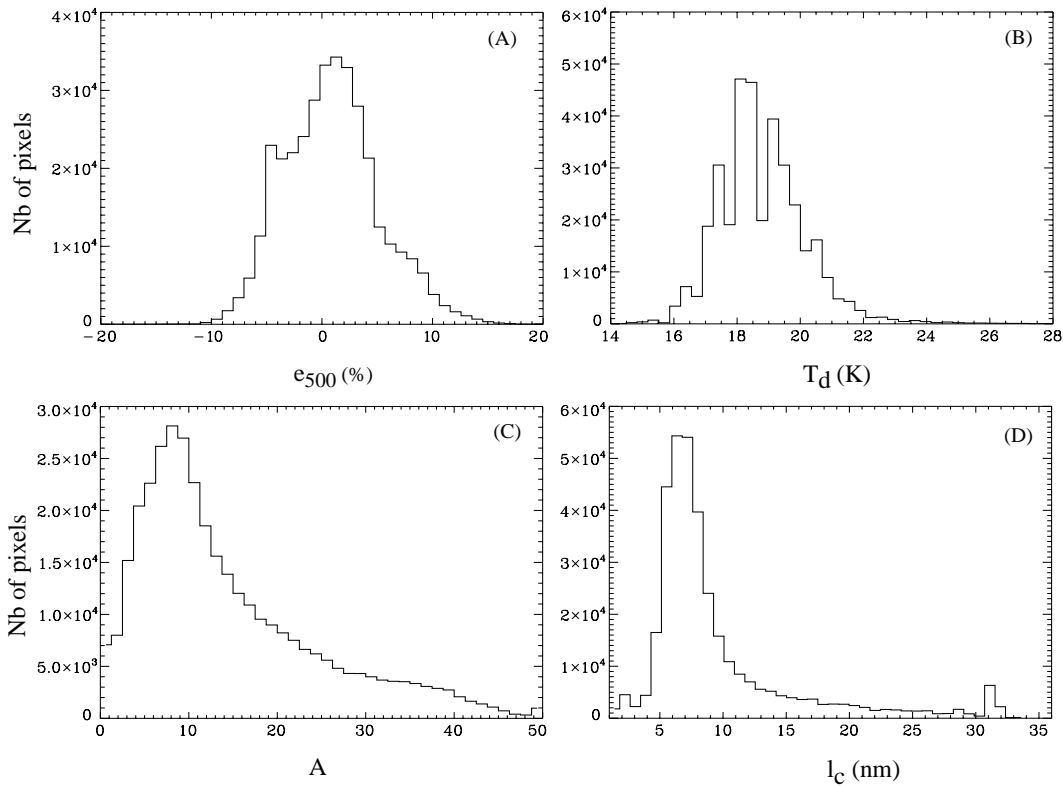}
\caption{Histograms of the 500 $\mic$ emissivity excess (panel A),
  results of the TLS modeling:
  the dust temperature (panel B), the $A$ parameter (panel C) as
  defined in equation \ref{eq_a}, and
  the $l_c$ parameter (panel D), as defined in equation \ref{eq_lc}.\label{fig_hist}}
\end{center}
\end{figure*}

Emissivity variations also appear
to be wavelength-dependent with the emission spectrum flattening in the
submillimeter ($\lambda >500-600$ $\mic$) relative to a single modified black-body emission
\citep{Wright91, Reach95, Finkbeiner99, Galliano05, Paladini07,
  Paradis09, Paradis11a}. They also appear to be environment-dependent
\citep[see, for instance,][for a review]{Li05},
with an absolute emissivity value in the FIR ($\lambda < 500-600$
$\mic$) that is higher in colder and denser environments \citep{Stepnik03,
  Paradis09, Abergel11}. Moreover, the analysis
of the Planck data in the solar neighborhood and both the Large and Small
Magellanic Clouds \citep{Bernard11a,Bernard11b} indicates that the
flattening of the emissivity spectrum could be related to
the metallicity, because it is more significant in low metallicity
environments. From the astrophysical
point of view, grain aggregation in the core of dense clouds could, in
principle, explain the observed absolute FIR emissivity excess, if one assumes
that aggregates emit more FIR radiation than individual
grains. Nevertheless, this interpretation is only partially
satisfactory, since grain aggregation may affect only the absolute values and
not necessarily the shape of the emissivity, and the aforementioned excess has only
been reported for some Galactic molecular clouds, while in the
extragalactic context, such as in the Large Magellanic Cloud
\citep{Paradis11b}, it has not been found at all.

In this work, we study the FIR/submm data obtained for the first and
fourth Galactic quadrants through the Herschel infrared GALactic
(Hi-GAL) survey \citep{Molinari10a, Molinari10b}, and demonstrate the
existence of a submm emissivity excess. Thanks to the sub-arcmin
resolution, comprehensive sky coverage, and broad wavelength range
spanned by the Hi-GAL data, we can for the first time perform a
systematic investigation of such an excess in our Galaxy. In addition,
the Galactic plane is ideal for conducting such an analysis because
it contains a variety of different environments, from diffuse clouds to compact sources, and from cold to warm structures, which can potentially harbor different dust properties. We also show that we are able to reconcile the observed excess  with theory predictions using the two level system (TLS) model, which provides a sophisticated analytical description of the disordered internal structure of amorphous dust grains.

The paper is organized as follows. In Section \ref{sec_data}, we
introduce the observational data that we use in this work. Sections
\ref{sec_excess} and \ref{sec_model} focus on the variations in the
emissivity spectral shape and the origin of these variations. Section
\ref{sec_cl} provides our conclusions.
 \begin{figure*}[!t]
\begin{center}
\includegraphics[angle=90]{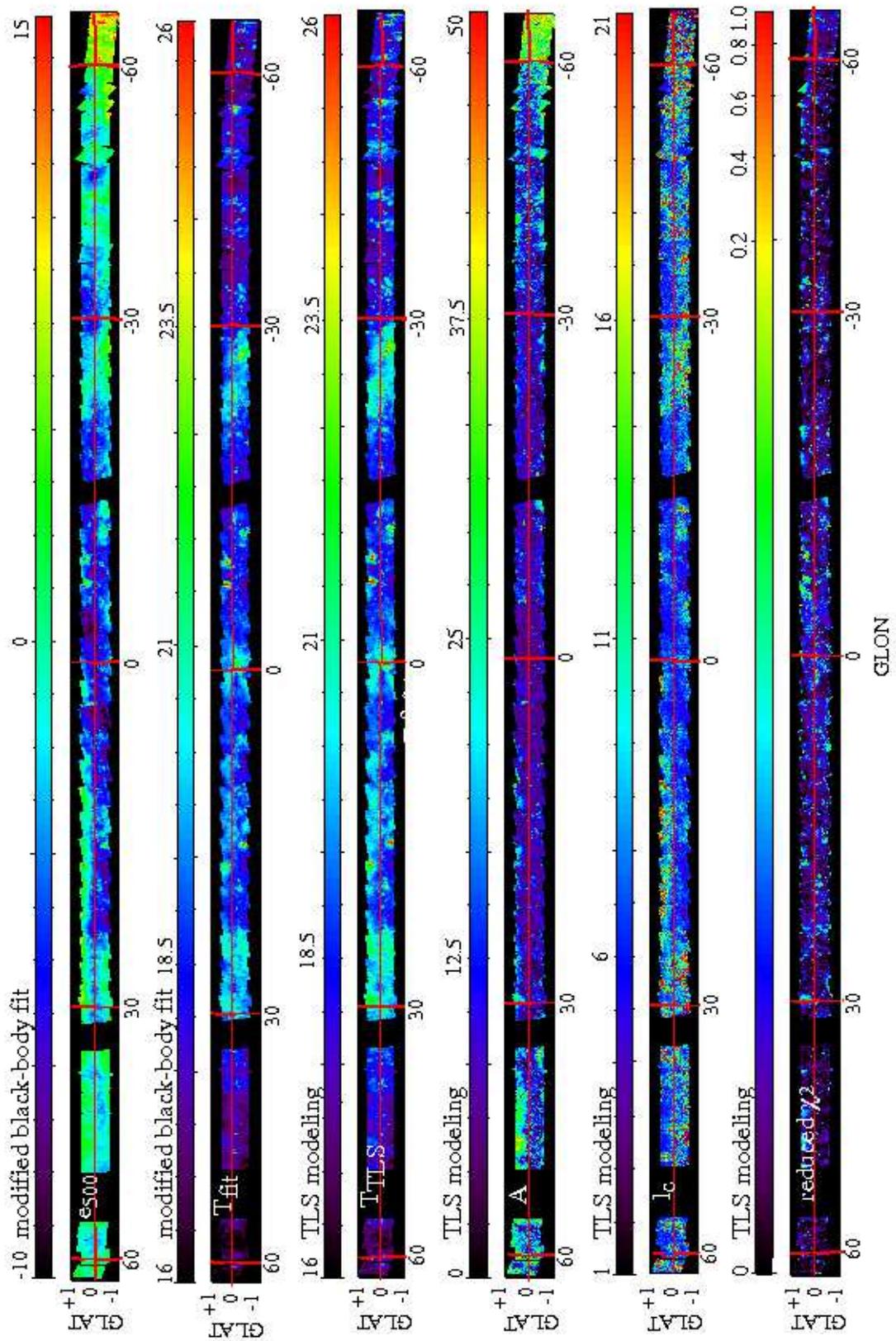}
\caption{From left to right: 500 $\mic$ emissivity excess (in percent)
  and dust temperature derived from a
  modified black-body emission, and the results of the TLS modeling for dust
  temperature ($T_{TLS}$) in K, intensity of the TLS processes ($A$), correlation length ($l_c$) in
  nm, and reduced $\chi^2$. \label{fig_excess}}
\end{center}
\end{figure*}
\begin{figure*}[!t]
\begin{center}
\includegraphics[width=14cm]{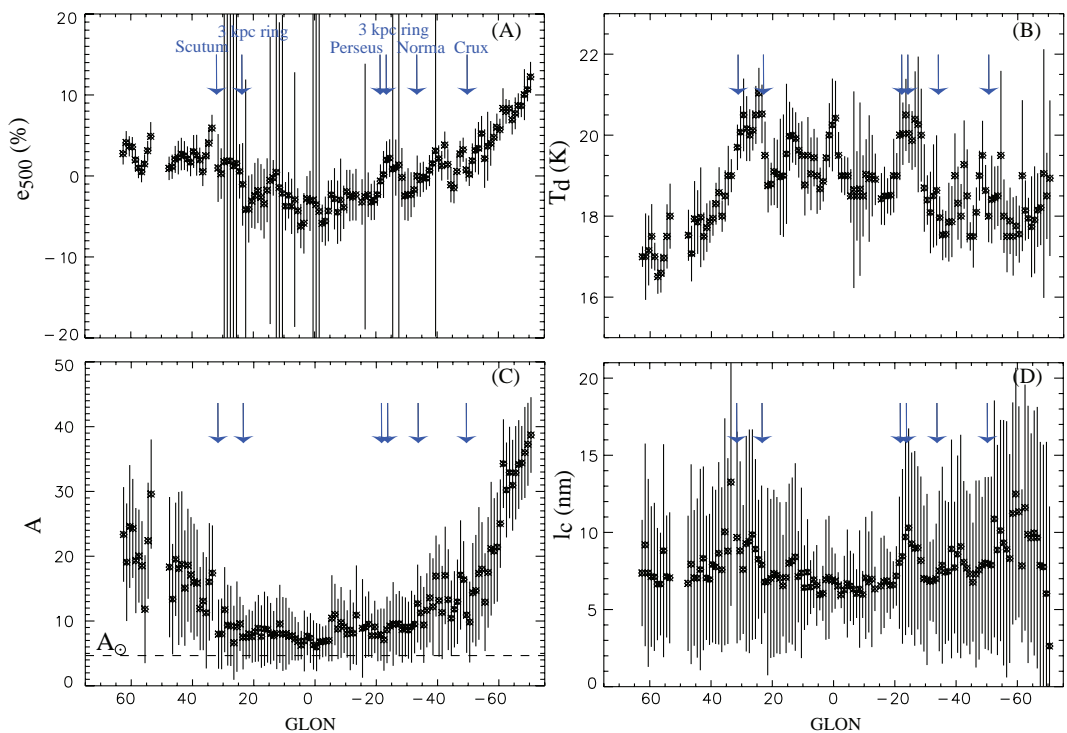}
\caption{500 $\mic$ emissivity excess (panel A), dust temperature
  (panel B), $A$ parameter (panel C), and $l_c$ parameter (panel D),
  as a function of Galactic longitudes. The dashed line shows the
  solar neighborhood value when normalized to $c_{\Delta,s}$=475. The vertical error
    bars correspond to the 1-$\sigma$ standard deviation. 
    The blue arrows show the tangent directions of the spiral arms at
    l=-50$\degr$, -33$\degr$, -21$\degr$, and 31$\degr$ \citep{Vallee08} and of the
    3 kpc ring at $|l|$=23$\degr$ \citep{Dame08}. \label{fig_long}}
\end{center}
\end{figure*}

\section{Data} \label{sec_data}
The reference database for our investigation consists of the
Hi-GAL dataset, a collection of 56 tiles of 2$\degr \times$ 2$\degr$,
with the longitude range -70$^{\circ} <$ l $<$ 70$^{\circ}$, and an
extension of 2$^{\circ}$ in latitude, following 
the Galactic warp. The Hi-GAL PACS (70 and 160 $\mic$) and
SPIRE (250, 350, and 500 $\mic$) data
\citep{Pilbratt10, Poglitsch10, Griffin10} were processed using the ROMAGAL software, described in
\citet{Traficante11}. We also used ancillary IRIS \citep[Improved Reprocessing of the IRAS Survey,
see][]{MamD05} data to constrain the peak of the BG
emission. We corrected the gains and zero-level in the SPIRE and PACS images by
applying gains and offsets derived from the
comparison to a combination of the Planck-HFI (High Frequency
Instrument) and the IRIS observations, using the
same procedure as in \citet{Bernard10}. A forthcoming paper will be
dedicated to the intercalibration of the Hi-GAL data with existing
photometric data \citep{Bernard11}. A full description of the
Planck mission is provided in \citet{Tauber10} and the \citet{Planckoverview}. The
Hi-GAL data ($\theta=6^{\prime \prime}$ to 37$^{\prime \prime}$ from
160 to 500 $\mic$) were convolved to a 4$^{\prime}$ angular resolution to match the resolution of the IRIS data. In addition, each tile was re-projected using the HEALPIx pixelization scheme
(Hierarchical Equal Area isoLatitude Pixelization)\footnote{http://lambda.gsfc.nasa.gov/} with 
nside=2048, which corresponds to a pixel size of 1.7$^{\prime}$,
i.e. adequate for Shannon sampling of the adopted resolution.
We used the surface area of the intersection between the HEALPix and the native pixel as a weight to reproject the data. After reprojecting, all the fields
were combined into a single HEALPix file covering the 
entire inner Galactic plane. We adopted the same absolute calibration
accuracy as for the SDP fields in the PACS data, i. e. 20$\%$,
and 13.5$\%$  for IRIS 100 $\mic$. The calibration of the SPIRE data
was recently revised leading to an accuracy of 7$\%$. As a
conservative approach, we did not include the PACS 70 $\mic$ data in
the analysis because at this wavelength we are probably most sensitive
to emission from
very small grains (VSGs), which are stochastically heated by photon
absorption \citep{Compiegne10}. This contribution can only be roughly evaluated and can severely bias 
the derivation of the BG temperature (especially for $T_d < 25$ K). 

\section{Emissivity excess at 500 $\mic$} \label{sec_excess}

A break in the observed emissivity law around 500 $\mic$ has been
identified in the outer Galaxy \citep{Paradis09} from the analysis of
DIRBE \citep{Hauser93}, Archeops
\citep{Benoit02}, and WMAP \citep{Bennett03}
data, as well as in the Large Magellanic Cloud based on
Herschel data \citep{Gordon10, Galliano11}. However, in the two Hi-GAL Science
Demonstration Phase (SDP) fields (centered at l=30$\degr$ and l=59$\degr$), 
\citet{Paradis10} did not find any significant emissivity
excess at 500 $\mic$, even when taking calibration uncertainties into account. 
The goal of the present work is to investigate the existence of this break. 
For this purpose, we fit the Hi-GAL/IRIS emission in the range 100 $\mic$ $< \lambda <$ 350 $\mic$ in each pixel with 
a single modified black body, adopting a least squares fit
  method. We assume that calibration uncertainties are distributed
  following a Gaussian function. The dust temperature is set as a free parameter. We then compare the observed emissivity at 500 $\mic$ ($\epsilon_{500}^{obs}$) with the value predicted by
extrapolating the best-fit emissivity power-law to 500 $\mic$
($\epsilon_{500}^{fit}$). We define the {\em{500 $\mic$ excess}} as the 
quantity 
\begin{equation}
e_{500}=\frac{\epsilon_{500}^{obs}-\epsilon_{500}^{fit}}{\epsilon_{500}^{obs}}.
\end{equation}
To perform the fit described above, we need to adopt a value for the emissivity spectral index $\beta$. 
A recent analysis of the Planck data \citep{Bernard11a} has found,
for 100 $\mic$ $< \lambda <$ 850 $\mic$, a median $\beta$ value of 1.8
over the entire sky. In contrast, 
\citet{Paradis10} obtained, for 100 $\mic$ $< \lambda <$ 500
$\mic$, a median $\beta$ value of 2.3 in both of the Hi-GAL SDP fields. In the
light of these results, we decide to keep $\beta$ fixed and equal to
2. We note that this assumption is a
matter of how the excess is defined and only affects the amplitude
of the excess and not its spatial distribution.\\

In Figure \ref{fig_hist} (panel A), we show the pixel distribution of the emissivity 
excess at 500 $\mic$ ($e_{500}$) derived from the fitting
procedure. The distribution is clearly not centered  at zero (as would
be the case for pure noise), but instead visibly skewed towards positive values. 
The median of the distribution is 0.8$\%$, with a $\sigma$ of 4.2$\%$. 
Excess values larger than
13.4$\%$ exceed the 3-$\sigma$ standard deviation of the distribution
and are significantly larger than the calibration uncertainty of the SPIRE data
(cf. Section \ref{sec_data}).  

We emphasize that this result, i.e. the existence of
a significant excess of emission at  500 $\mic$, does not depend on the
particular value adopted for $\beta$ in the modified black-body fit. 
If we repeated the analysis with a different spectral emissivity value, for example 
$\beta$ equal to 1.8 instead of 2, the shape of the excess distribution would remain unchanged, although it would shift towards more negative 
values, suggesting that, for the majority of the 
pixels, the model overestimates the data. Incidentally, we note that
this also corroborates the hypothesis that, at least along the
inner Galactic plane, $\beta$ is indeed larger than 1.8. 
\begin{figure}[!t]
\begin{center}
\includegraphics[width=8.5cm]{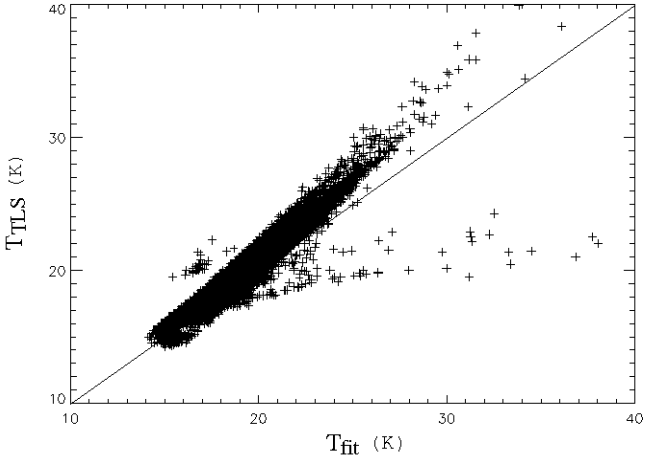}
\caption{Temperature derived using the TLS model ($T_{TLS}$), as a
  function of temperature derived from a modified black-body fit with
  $\beta$=2 ($T_{fit}$).\label{fig_compare_t}}
\end{center}
\end{figure}

Figure \ref{fig_excess} illustrates the spatial distribution of the
500 $\mic$ excess across the Galactic plane. Inspection of the map reveals a more 
pronounced excess in the peripheral regions (($35\degr<|l|<70\degr$))
than in the central ones ($|l|<35\degr$). This effect also appears in Figure \ref{fig_long}, panel A, where we have plotted the median value of the 
excess at each Galactic longitude, and its associated 1-$\sigma$ standard
deviation. In particular, for $-70\degr<l<-60\degr$
the excess becomes significant, contributing to as much as 16 - 20$\%$ of the total 
emissivity, and even reaching values larger than 23$\%$ toward
some HII regions. Regions in the proximity of the Galactic center, where a ring of
dense and cold clouds has been identified at the Herschel
resolution \citep{Molinari11}, however, do not exhibit any
excess. The spatial distribution of the excess does not display any
particular symmetry either: for instance, the significant increase in the fourth Galactic quadrant  
around l$<$-50$\degr$ is not mirrored by a corresponding increase
within the same longitude range in the first quadrant. Remarkably, the morphological 
behavior that characterizes the 500 $\mic$ excess is - per se - an indication that 
this effect is not related to calibration uncertanties, as these would
have an equal impact 
on different regions of the plane. In the specific fields of l=30$\degr$
and l=59$\degr$, the median value of the excess is 1$\%$ and 3$\%$,
which does not correspond to any significant detection, in agreement
with the findings in \citet{Paradis10}.

The submm excess could, in principle, be induced by the fitting
procedure. To test this hypothesis, we carried out two 
tests: first, we included in the fit all data points for 100 $\mic$ $\leq$ $\lambda$ $\leq$ 
500 $\mic$; second, we performed the fit by replacing the 350 $\mic$
data point with the 500 $\mic$ data. Both tests provided larger reduced $\chi^2$ than the
original fit (on average, by a factor of 2.3 and 2.6), thus supporting the idea that this excess has an  
astrophysical origin. 

In the following, we investigate in detail the 500 $\mic$ excess in
the framework of the TLS model, by analyzing variations in the model parameters.

\section{Towards an understanding of the nature of the excess} \label{sec_model}

\subsection{The TLS model}
We now provide a brief description of the TLS model. We
refer the reader to \citet{Meny07} for a fuller description. 

Our revised TLS model is based on both solid
state physics and laboratory measurements, and provides an accurate description 
of the physical properties of amorphous grains. It consists of two  
parts: (1) the disordered charge distribution (DCD) one, which describes the interaction between
the electromagnetic wave and acoustic oscillations in the disordered
charge of the amorphous material \citep{Vinogradov60,Schlomann64}; and 
(2) the actual TLS part, which takes the interaction of
the electromagnetic wave with the simple distribution of an asymmetric
double-well potential into account. This second part corresponds to a 
theory originally developped by \citet{Phillips72, Phillips87} and  \citet{Anderson72}. 
The first effect (DCD) is temperature-independent, occurs on
the grain scale and is the most dominant effect in the FIR. 
Moreover, it has two types of asymptotic behaviors at both lower and
higher frequencies than $\omega_0$: $\epsilon_{\nu} \propto \lambda^{-2}$ in the short
wavelength range, i. e. $\omega> \omega_0$, and $\epsilon_{\nu}
\propto \lambda^{-4}$ in the long wavelength range, i. e $\omega<
\omega_0$. This DCD effect is characterized by the so-called {\em{correlation length}} 
l$_{c}$, which determines the wavelength at which the inflection point
between the two asymptotic behaviors occurs. This parameter is defined
in the $\omega_0$ term by
\begin{equation}
\label{eq_lc}
\omega_0=v_t/l_c,
\end{equation}
where $\rm v_t$ is the transverse sound velocity in the material.
The TLS effects, which consist of three mechanisms (a resonant absorption and two
relaxation processes, thermally activated), are instead temperature-dependent, take place 
on the atomic scale, and start to be important in the submm domain,
becoming the predominant effects in the mm wavelength range. The
  amplitude of the TLS effects with respect to the DCD part is
  determined by the intensity parameter denoted with $A$. This means
  that all TLS processes are multiplied by the same intensity $A$. The total
  emission deduced from the TLS model ($I_{tot}$) is then the sum of the emissions
  coming from the DCD ($I_{DCD}$) and TLS processes ($I_{TLS}$), such
  as
\begin{equation}
\label{eq_a}
I_{tot}=I_{DCD}+\sum I_{TLS}.
\end{equation}

\begin{figure}[!t]
\begin{center}
\includegraphics[width=8.5cm]{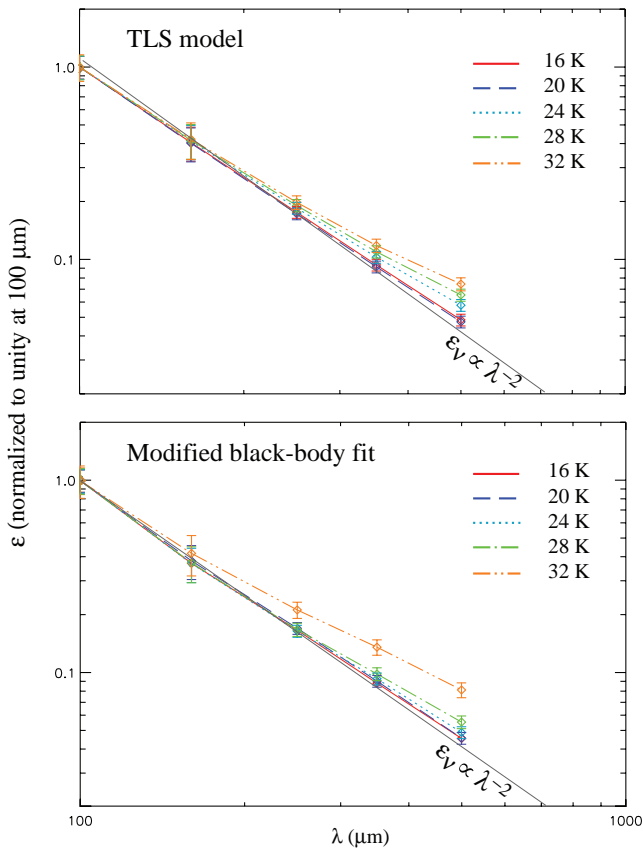}
\caption{Average emissivity spectra in five bins of temperature and their
  1-$\sigma$ uncertainty: 16 K in red (solid), 20 K in dark blue
  (dash), 24 K in light blue (dot), 28 K in green (dash-dot), and 32 K
  in orange (dash-dot-dot). The emissivities have been computed
  using the best-fit with the TLS model (top panel)
  and with a modified black body with $\beta$=2 (bottom panel).An
  emissivity power-law in $\lambda^{-2}$ is overplotted as a continuous
  gray line, for comparison. \label{fig_em}}
\end{center}
\end{figure}

\citet{Paradis11a} performed a comparison
between the model predictions and astrophysical data to determine the model
parameters. They analyzed both the FIR/mm SED of
the solar neighborhood using FIRAS and WMAP
data, and of Galactic compact sources observed with the Archeops
experiment. The purpose of their analysis was to constrain the four
key parameters of the TLS model, namely: (1) the charge correlation length $l_c$ for the DCD absorption;
(2) the amplitude $A$ of the TLS effects, with respect to the DCD
process; (3) a parameter $c_{\Delta}$ describing the double-well
potential for one of the TLS processes; and (4) the dust temperature
$T_{TLS}$. The combined fit of the SEDs for the solar neighborhood and the compact sources 
provided a general description of Galactic dust in terms 
of these parameters, whose best-fit values are $l_{c,s}$=13.4 nm, $A_s$=5.8, and
$c_{\Delta,s}$=475 (see also Table \ref{tab_val_med}), which were referred to as
the {\em{standard}} values.

In the following, we repeat the fitting procedure
described in Section \ref{sec_excess}, using 
this time the TLS model rather than a modified black body. The
best-fit values for the TLS parameters derived from the fit are then
compared to the standard ones, and those obtained for the solar
neighborhood and compact sources. Before describing the results we obtained, a few considerations need to be made. First, we note
that the $c_{\Delta}$ parameter is, to first order, degenerate with
the $A$ parameter. In addition, $c_{\Delta}$ cannot be tightly constrained in our study because 
we are limited to wavelengths shorter than 500 $\mic$. For this reason, following 
\citet{Paradis11a}, we set $c_{\Delta}$ to be equal to the standard value, ($c_{\Delta,s}$ = 475).

As for $l_c$, we adopt 36 nm as an upper limit. For $l_c \geq
36$ nm, the DCD process indeed reaches an asymptotic behavior in
$\lambda^{-2}$ from FIR to mm wavelengths. Values of $l_c$ of the
order of a few nanometers, combined with either small or null values of $A$, allow $\beta$ to be instead 
close to 4. The slope of the spectra increases with decreasing
$l_c$, in the absence of significant TLS processes. As analyzed in
\citet{Paradis11a}, for the range 100-550 $\mic$, $A$=0.1 and
$l_c$=1 nm (30 nm), the parameter $\beta$ is expected to be constant with a value
of 3.4 (2.0), regardless of the dust
temperature (see their Figure 7, middle panel). However, for $A$=10
and $l_c$=1 nm (30 nm), $\beta$ varies from 2.45 ($\simeq$1.85) to
1.4 (1.35) for temperatures between 10 K and 55 K. 

\subsection{Dust properties along the Galactic plane}
\subsubsection{Method}
To minimize the computing time, we pre-calculated the brightness
in the IRIS 100 $\mic$ and Herschel bands using the TLS model, taking
into account the color-correction to be applied to each instrument. In
this way, we generated a multi-dimensional grid, for different $T_{TLS}$ between
10 K and 40 K (with a 0.5 K step), $l_c$ in the range 1-36 nm (with a 1 nm step), and 50 values
of A between 0 and 50. We then compared the 
brightness expected from the TLS model with the observed brightness in each
pixel of the map. The $\chi^2$ value was computed for each value of the grid, and we chose 
the value of the parameters ($p$) that minimizes the $\chi^2$. Calibration
uncertainties were included in the fit as described in Section
\ref{sec_excess}. To allow interpolation between
individual entries of the table, the best-fit parameter values ($p^{\star}$) are
computed for the ten smallest values of $\chi^2$ using
\begin{equation}
p^{\star}= \frac{ \sum_{i=1}^{10} p_i\times
  \frac{1}{\chi^2_i}}{\sum_{i=1}^{10} \frac{1}{\chi^2_i}}.
\end{equation}
Since Planck data are not yet available, we are unable to place tight constraints  
on the $A$ parameter using only Herschel data. 

A normalization of 
the dust emission predicted by the TLS model is required. We adopt the normalization
\begin{equation}
I_{TLS,norm}=\frac{I_{TLS}\times <I_{obs}>}{<I_{TLS}>},
\end{equation}
where $I_{TLS}$ and $I_{TLS,norm}$ are the emission spectrum from the TLS model 
before and after the normalization, while $<I_{obs}>$ and $<I_{TLS}>$ are the
emission averaged in wavelength between 100 $\mic$ and 500 $\mic$, corresponding, respectively, to the observations and the TLS model. 

\begin{figure*}[!t]
\begin{center}
\includegraphics[angle=90]{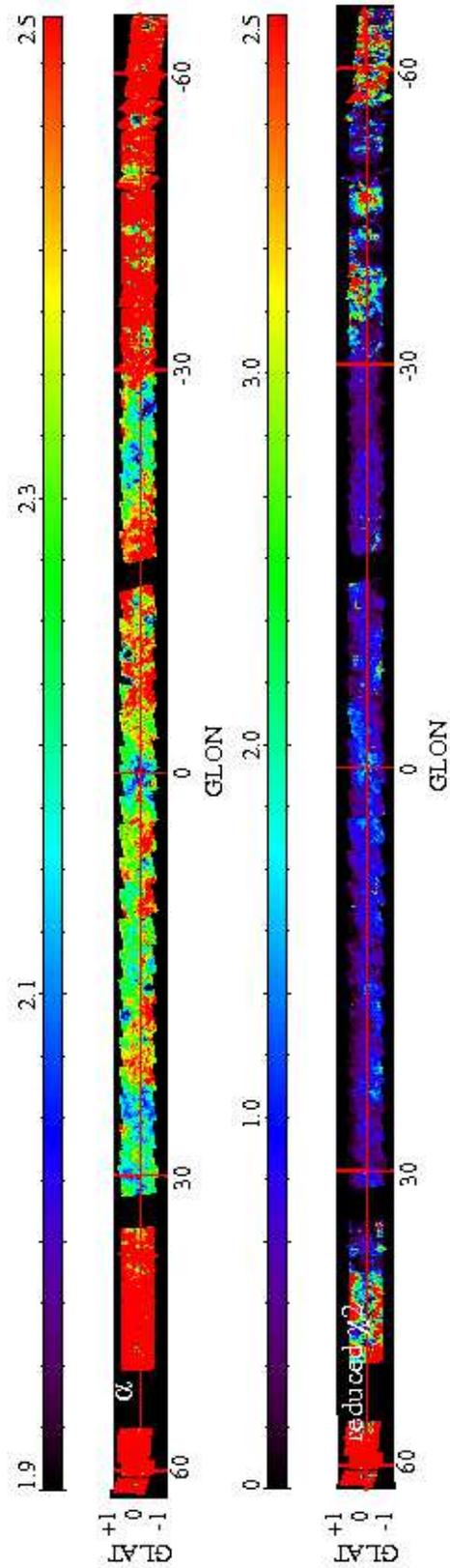}
\caption{Left panel: $\alpha$ parameter from the \citet{Dale01}
  model. Right panel: reduced $\chi^2$ deduced from the comparison
  between data and the \citet{Dale01} model. \label{fig_dale}}
\end{center}
\end{figure*}

\subsubsection{Dust temperature and emissivity}
\label{sec_t}
The spatial distribution of the dust temperature, as well as the
reduced $\chi^2$ obtained through the TLS modeling are presented in Figure
\ref{fig_excess}. The temperature histogram is shown in Figure
\ref{fig_hist} (panel B). 
We note that the reduced $\chi^2$ values are systematically smaller
than 1, 
suggesting that the uncertainties in the observational data are
probably overestimated, particularly for the PACS data. The median temperature along the Galactic plane is 18.6 K, while the Galactic
center exhibits temperatures in the range 18-25 K. As reported in
\citet{Molinari11}, some cold clumps can be found in this region, with 
temperatures of about 16-18 K. However, with respect to \citet{Molinari11}, we do not 
find any evidence of temperatures above 25 K. This may be for several reasons.
First, our temperature map was generated at 4$^{\prime}$ angular
resolution, instead of 25$^{\prime \prime}$ as in
\citet{Molinari11}. Secondly, the 70 $\mic$ was not included in
our analysis, which might have caused the underestimation of the warmest dust temperatures.

In general, temperatures appear to be higher in the central regions of
the Galaxy ($T_{TLS}$=19.4 K for $|l|<35\degr$), 
and to decrease with increasing longitude ($T_{TLS}$=18.0 K for
$35\degr<|l|<70\degr$), as one can clearly see in
Figure \ref{fig_long} (panel B). This general behavior has been seen before
using DIRBE \citep{Sodroski94}, FIRAS \citep{Reach95},
and Planck data \citep{Bernard11a}. Temperatures span the range 16 K - 26 K (Figure \ref{fig_hist},
panel B). A the IRIS resolution, the Galactic plane remarkably does
not show dust temperatures below 15 K. Some warm spots with $T_{TLS}\geq$ 30 K are, however, 
visible in the peripheral regions, probably owing to HII
regions. These warm spots also show the largest 500 $\mic$ emissivity excess. The
comparison between temperatures estimated with the TLS model
and those derived from the modified black-body fit in the range 100-350 $\mic$ with
$\beta$=2 ($T_{fit}$) is provided in Figure \ref{fig_compare_t}. As described
in \citet{Paradis11a}, there is good agreement between the two
methods up to $\simeq$25 K, where a departure from a one-to-one
correlation becomes noticeable. The dust temperature is only taken
into account in the Planck function in the case of a modified black-body
fit, whereas the TLS processes are temperature-dependent and increase with 
temperature.  As a consequence, the difference in the temperature contribution
in the two methods could explain the departure in the $\rm T_{TLS}$-$\rm
T_{fit}$ correlation at high temperatures, and especially the faster
increase in $\rm T_{TLS}$ compared to $T_{fit}$ for temperatures higher
than 25 K. The outlier points ($T_{fit}>20$ K) correspond either to
pixels with a low quality $\chi^2$ for the modified black-body fit, or to
edges of the map.\\

In Figure \ref{fig_em}, we have plotted the average dust emissivities,
for 100 $\mic$ $< \lambda <$ 500 $\mic$, in five dust temperature bins
centered at 16, 20, 24, 28, and 32 K. The plots have been 
created for both the TLS model and a modified black-body fit with
$\beta$=2. In the first case (TLS model), we computed the dust emissivity
  spectra using the best-fit model parameters ($T_{TLS}$, $A$, and $l_c$) for each pixel, that we
  averaged per temperature bin. We derived the averaged emissivity
  in the photometric bands by applying the 
  adequate color-correction for each instrument. In the second case (modified
black-body fit), we divided the brightness coming from the data by the
Planck function at the temperature $T_{fit}$. The emissivity spectra
were normalized to unity at 100 $\mic$. 
The uncertainties in the emissivity were computed using
\begin{equation}
\frac{\Delta
  \epsilon_{\nu}}{\epsilon_{\nu}}=\frac{\Delta
  B_{\nu}(T_d)}{B_{\nu}(T_d)}+\frac{\Delta I_{\nu}}{I_{\nu}}.
\end{equation}
The relative error $\Delta B_{\nu}(T_d)/B_{\nu}(T_d)$
is related to the error on $T_d$ ($\Delta T_d$) as
\begin{equation}
\frac{\Delta
  B_{\nu}(T_d)}{B_{\nu}(T_d)}=\frac{\frac{h\nu}{kT_d}e^{\frac{h\nu}{kT_d}}}{e^{\frac{h\nu}{kT_d}}-1}
\frac{\Delta T_d}{T_d}
\end{equation}

The figure highlights the existence, for both models, of a clear trend in the emissivity
spectra with the dust temperature, with the spectra being flatter
with increasing dust temperature. This observed flattening behavior with $T_d$ cannot be 
accounted for by the calibration error at 500 $\mic$. If the brightness at 500 $\mic$ were
systematically off by 7$\%$ owing to calibration uncertainties, all
emissivity spectra would be affected but this should not induce a change
in the dust temperature.
Another effect has to be considered. Given that we conducted our
analysis along the Galactic plane, where temperature mixing is expected along the LOS, one may argue that we
can introduce an artificial bias by fitting the data points with a single dust temperature. To address this question, following \citet{Dale01},
we computed the predicted brightness
in each IRIS and Herschel band for different values of the
interstellar radiation field intensity, $X_{ISRF}$, relative to
  the intensity of the ultraviolet field in the solar neighborhood \citep{Mathis83}.

The \citet{Dale01} model describes a temperature distribution using
the concept of local SED combination, and assumes a power-law
distribution of dust mass subjected to a given heating intensity $dM_d(X_{ISRF})$
\begin{equation}
dM_d(X_{ISRF}) \propto X_{ISRF}^{\alpha}dX_{ISRF},
\end{equation} 
where $\alpha$ and $X_{ISRF}$ are in the range 1$<\alpha<$ 2.5 and
0.3$<X_{ISRF}<10^5$. With this representation, the diffuse medium has 
$\alpha$ close to 2.5, whereas active star-forming regions have $\alpha$ of 1. 
We estimated the emission spectra ($I_{\nu}^{mod}(X_{ISRF})$), for different values of $\alpha$, using the DustEM
package \citep{Compiegne08, Compiegne11} and assuming $\beta$=2. We then summed these contributions over the same $X_{ISRF}$ range proposed by \citet{Dale01}
\begin{equation}
I_{\nu}^{tot}=\frac{\sum_{i=1}^{88} \sum_{j=1}^{50} I_{\nu}^{mod}(X_{ISRF,i}) \times X_{ISRF,i}^{-\alpha_j}}{\sum_{i=1}^{88} \sum_{j=1}^{50} X_{ISRF,i}^{-\alpha_j}}.
\end{equation}  
We then compared the observational data with the predicted SEDs, and searched for the optimal
$\alpha$ value that minimizes the difference between the
two. As a result of this process, each pixel of the map has an
emission that incorporates temperature mixing along the LOS. Figure
\ref{fig_dale} is an illustration of the pixel-to-pixel map of the
recovered $\alpha$ values. The central regions of the Galactic plane are 
characterized by values of $\alpha$ in the range [1.9, 2.5], while for
$|l| >$ 30$^{\circ}$, $\alpha$ appears to be systematically equal to
the asymptotic value of 2.5. At the same type, inspection of the reduced
$\chi^2$ spatial distribution (also shown in Figure \ref{fig_dale})
reveals that the SEDs for most of the peripheral regions are not well
reproduced by the \citet{Dale01} model. In these regions, temperature mixing is probably less
important than in the central regions of the Galaxy. 

We considered only pixels with reduced $\chi^2<1.5$. Modeling with a reduced $\chi^2 \geq 1.5$ does
not give a satisfactory fit. The spatial distribution of the reduced
$\chi^2$ is provided in Figure \ref{fig_dale}. This selection removes 39$\%$ of the
pixels, and in particular the peripheral parts or the map, corresponding essentially
to the coldest parts of our map. We then proceeded as before, with the difference that 
we treated the brightness derived from the best-fit \citet{Dale01} model for
the selected pixel (61$\%$ of the Galactic plane) as ``real'' data: we performed a
$\chi^2$ minimization, using the TLS model and a modified black-body
emission model with $\beta$=2, to derive the best-fit parameters and
the emissivity spectra for different temperature bins. Since the
coldest regions had been rejected from the analysis, the temperature bins ranged 
from 19.1 K to 31.9 K.

\begin{figure}[!t]
\begin{center}
\includegraphics[width=8.5cm]{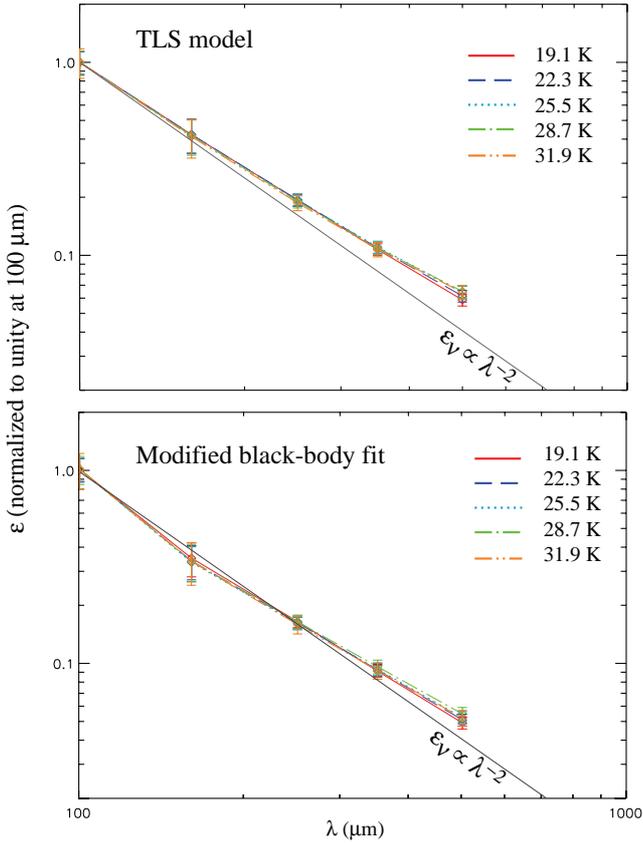}
\caption{Emissivity spectra in five bins of temperature and their
  1-$\sigma$ uncertainty, assuming
  temperature mixing along the LOS (with the use of the \citet{Dale01}
  model): 19.1 K in red (solid), 22.3 K in dark blue
  (dash), 25.5 K in light blue (dot), 28.7 K in green (dash-dot), and 31.9 K
  in orange (dash-dot-dot). The emissivities have been computed
  for the best-fit TLS model (top panel)
  and a modified black body with $\beta$=2 (bottom panel). An
  emissivity power-law in $\lambda^{-2}$ is overplotted for comparison in continuous
  gray line.\label{fig_em_mixt}}
\end{center}
\end{figure}

If temperature mixing did indeed account for the flattening of the spectra with increasing
dust temperature, we would expect to observe the same type of behavior
with this test. However, as shown in Figure \ref{fig_em_mixt}, the
spectra do not show any sign of a dependence on dust temperature, regardless of the model used to derive
the dust emissivities. Thus, temperature mixing is
 not likely to be responsible, and we are confident that the
emissivity changes are instead related to the intrinsic properties of dust grains.

\subsubsection{Model parameters: $A$ and $l_c$}
Unlike in \citet{Paradis11a}, we explored both cold and warm/hot regions of the Galaxy,
to systematically investigate the variation in the  TLS model parameters with environment. 

\begin{figure}[!t]
\begin{center}
\includegraphics[width=8.5cm]{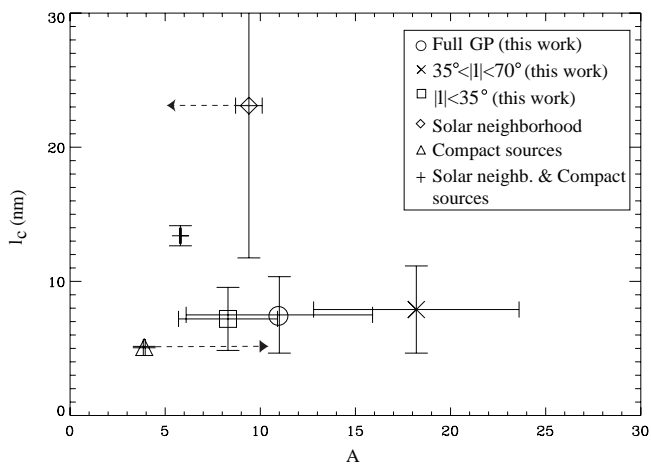}
\caption{Best-fit model parameters $l_c$ and $A$ and their 0.5-$\sigma$
  standard deviations, derived from the TLS model for different
  Galactic environments. The arrows
  indicate values of the parameter to first order, when fixing $c_{\Delta}$ to the
  standard value equal to 475. \label{fig_param}}
\end{center}
\end{figure}

Figure \ref{fig_param} shows the median values derived from this study
for the $l_c$ and $A$ parameters in the case of, respectively, the
whole Galactic plane region covered by Hi-GAL and, the
central and peripheral parts of the plane. We have also displayed, for comparison, the best-fit values deduced by
\citet{Paradis11a} for the solar neighborhood, for Archeops compact
sources, and for the combination of both. In this analysis, the uncertainties in the parameters are
assumed to be equal to their standard deviation. Since the error bars in the data are likely to be overestimated, as already discussed in Section \ref{sec_t}, the parameters cannot be strongly constrained.

The distribution of points in the figure reveals that different environments are likely associated with 
different dust grain properties. From the fitting procedure
discussed in Section 4.2.1, we estimated that the most
representative values of $l_c$ and $A$ for the Galactic plane are 7.5
nm and 11.0. For $l_c$, this is close to the value derived for
Archeops sources, but smaller than for the solar neighborhood. We also
note (Figure \ref{fig_hist}, panel D) that $l_c$ is globally
characterized by values much lower than the standard one ($l_{c,s}$ = 13.4
nm). Both of these results are indicative of a less organized charge distribution
for dust grains in the plane than for dust particles in the
solar neighborhood. As for the amplitude of the TLS processes, $A$, this
is larger by a factor of 1.9 in the Galactic plane than the standard
value ($A_s=5.8$). We note, however, that in the solar neighborhood, probed by
FIRAS observations, and in Archeops compact sources, $c_{\Delta}$ differs significantly from the standard value. Therefore, to be able to 
perform a proper comparison between these results, we needed to normalize $A$ to 
$c_{\Delta,s}$=475. Accordingly, we obtained $A$ equal to 4.8 and 10.9 for, respectively, 
the solar neighborhood and the compact sources. Again, $A$ in the Galactic
plane is different from the value derived in the solar
neighborhood. This result indicates that
grains within the plane might have a larger degree of
amorphization than in the solar neighborhood on the atomic scale, where the TLS processes
occur. 

\begin{figure}[!t]
\begin{center}
\includegraphics[width=9cm]{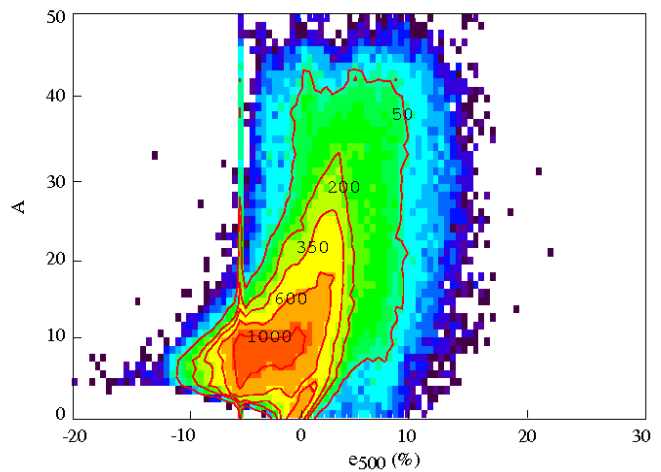}
\caption{$A$ parameter derived from
  the TLS model as a function of the 500 $\mic$ emissivity
  excess. The overplotted contours, with levels at 50, 200, 350, 600, and 1000, represent the density of pixels. \label{fig_exc_flux}}
\end{center}
\end{figure}

The spatial variation across the Galactic plane of $l_c$ and $A$ is provided in
Figure \ref{fig_excess}. Important variations in the two parameters, on
both small scales (pixel to pixel) and large scales (from the central
to the peripheral parts of the Galaxy), are visible. The $A$ parameter exhibits high values in
regions with significantly high $e_{500}$ (especially in peripheral regions,
see Figure \ref{fig_long}, panel C), which suggests that the excess does not depend on $l_c$, but 
rather on $A$. This explains the large fluctuations in the $l_c$
  parameter in the peripheral regions (see Figure \ref{fig_excess}), which
  induce large standard deviations in the longitude profile of this
  parameter (see Figure \ref{fig_long}). In contrast, $l_c$ is
  more accurately characterized in regions of negative excess. In that latter case, variations in the spectral shape of the SEDs
  can only be accounted for by variations in the correlation length,
  since this parameter allows us to produce emission spectra with a
  spectral index larger than 2. This explains why the $l_c$ standard
  deviation is smaller closer to the Galactic center. 
The correlation between the 500 $\mic$ excess and
the $A$ parameter is shown in Figure \ref{fig_exc_flux}. Disorder in
the grain structure on the atomic scale, i. e. the TLS part of the
model, thus appears to be the main
contributor to the emissivity variations at 500 $\mic$. The $l_c$
parameter is indeed almost constant along the Galactic plane, as one can
discern in Figure\ref{fig_long} (panel D). 

The $A$ values (or excess values) are higher over most of the inner
Galactic plane than in the solar neighborhood. If the origin of the
variations is a general Galactic gradient, this would indicate that
the solar neighborhood is an atypical place. It might also suggest
that the spatial distribution of the excess is not described by simply an overall
Galactic gradient but a more complicated structure.

The relation between $A$ and $l_c$ is illustrated in Figure
\ref{fig_A_lc}. The plots are very noisy, especially for the whole 
and the peripheral regions of the Galactic plane. Nevertheless, they reveal a weak anti-correlation between the two parameters, which disappears at large values of $A$. We also see
that, in the central regions of the Galactic plane, most of the points cluster in the region delimited by $A\simeq
6-10$, whereas the peripheral regions is rather accurately characterized by
$A\simeq10-22$. This shift in $A$ is likely associated with variations in
the dust properties, while the disappearance of the anti-correlation between $A$ and $l_c$
confirms that spectral variations at wavelengths equal to or larger 
than 500 $\mic$ could be entirely due to variations in the intensity of the TLS
process. A significant flattening of the spectra in the submm/mm domain is indeed expected if the TLS effects are responsible 
for the observed excess. We remind the reader, however, that
longer wavelength observations are required to confirm the results 
obtained for this parameter.
We note that the calibration uncertainty in the 500 $\mic$ measurement does not affect
the spatial distribution of the TLS parameters. The calibration error
would indeed only induce a variation in the absolute value of $A$. For instance, a decrease in the SPIRE
500 $\mic$ data by 7$\%$ would cause a corresponding decrease in $A$
by a factor of 1.6, but the peripheral regions of the plane would still have the largest values of $A$. As for $l_c$, its median
value - and standard deviation - would be lower ($l_c\simeq$5.8 nm)
than the regular median value ($l_c\simeq$7.5 nm),
even if the change in the dust temperature is as small as 0.1 K. The difference of 1.4 K between the average 
dust temperature in the central and peripheral regions of the Galaxy would be preserved (see Section \ref{sec_t}). 

\begin{table*}[!t]
\begin{center}
\begin{tabular}{ l  c  c  c c }
\hline
\hline
& $T_{TLS}$ (K) & $l_c$ (nm)& $A$ & $c_{\Delta}$\\
\hline
Full Hi-GAL data & 18.6$\pm$1.4 & 7.5$\pm$5.7 & 11.0$\pm$9.8 & 475$^*$ \\
Peripheral Hi-GAL data ($35\degr<|l|<70\degr$) & 18.0$\pm$1.2 & 7.9$\pm$6.5 & 18.2$\pm$10.8  & 475$^*$ \\
Central Hi-GAL data ($|l|<35\degr$) & 19.4$\pm$1.2 & 7.2$\pm$4.7 & 8.3$\pm$5.2 & 475$^*$ \\
Solar neighborhood & 17.5$\pm$0.02 & 23.1$\pm$22.7 & 9.4$\pm$1.4 &
242$\pm$123\\
& & & 4.8$^{**}$&475$^{*}$ \\
Compact sources & - & 5.1$\pm$0.1 & 3.9$\pm$0.1 & 1333$\pm$68\\
& & & 10.9$^{**}$&475$^{*}$ \\
Solar neighborhood + Compact sources & 17.3$\pm$0.02& 13.4$\pm$1.5 & 5.8$\pm$0.1
& 475$\pm$20 \\
\hline
\end{tabular}
\caption{\label{tab_val_med} Parameters of the TLS model and their 1-$\sigma$ standard deviation: median values of the best-fit
  parameters for the Galactic
  plane (this work) and best-fit parameters for the other environments
  \citep[from][]{Paradis11a}.\newline
$^*$ set to the standard value defined in \citet{Paradis11a}.\newline
$^{**}$$A$ normalized to $c_{\Delta,s}=475$.}
\end{center}
\end{table*}

\subsubsection{Relationship with the Galactic spiral arms}
According to \citet{Greenberg95}, the spiral arms regions and the
interarm regions might have different dust properties. This could explain
the variations with Galactic longitude observed in the extinction data 
of \citet{Gao09}. The latter authors argued that these variations could
be caused by the larger grains produced by coagulation, a higher dust density
and a stronger radiation field in the spiral arms regions.
\citet{Bernard10} also showed the existence of warmer dust in regions of
the LOS corresponding to the intersection with spiral arms. In Figure
\ref{fig_long}, we have marked the tangent positions of the spiral
arms taken from \citet{Vallee08} (l=-50$\degr$, -33$\degr$, -21$\degr$,
and 31$\degr$ for the Crux-Scutum, Norma, Perseus, and Scutum-Crux arms, respectively), as well as the position of the 3-kpc
molecular ring defined in \citet{Dame08} ($|l|=23\degr$).  

From Figure \ref{fig_long}, one can observe a direct link between the
tangent positions of the spiral arms with some peaks in the dust
temperature profile. Variations in the 500 $\mic$ emissivity excess
linked to the arms are also visible but are less significant than the
general trend with distance from the Galactic center. For the $A$
parameter, the relationship with spiral arms is less obvious. A hint
of a relationship can be discerned between the spiral
arms and the correlation length, which is enhanced in these
regions, although by an amount that is smaller than the standard deviation.

Variations in the optical properties predicted by the TLS model replace or
are complementary to the predicted variations in emissivity caused by
either aggregation or fluffiness. Dust growth is expected in dense molecular
clouds, and aggregation likely produces both an increase in emissivity
with respect to the gas column (for a given dust to gas ratio) and a
decrease in dust temperature. The increase in the absolute emissivity with
respect to the gas cannot be measured here because we do not
directly compare the emissivity with the gas column density. From dipole discrete approximations
calculations, the emissivity increase for aggregates is expected to be roughly
constant in the range 100-500 $\mic$. This does not change the
spectral shape of the emissivity and cannot reproduce any significant
500 $\mic$ emissivity excess. Similarly, the changes in apparent dust
temperatures are difficult to interpret in terms of aggregation at the
scales analyzed here, and are probably determined mainly by variations
in the local radiation field intensity, which may obscure any temperature
decrease caused by dust aggregates. 

\section{Conclusions} \label{sec_cl}
We have performed an analysis of the emissivity variations along the Galactic plane 
using the newly released Herschel/Hi-GAL data (160 $\mic$ $< \lambda <$ 500 $\mic$) 
combined with the IRIS 100 $\mic$ data, at 4$^{\prime}$ angular
resolution. Changes in the emissivity spectra are interpreted in terms
of the TLS model, which includes the DCD process characterized by a
correlation length ($l_c$), and the TLS effects, whose intensity ($A$) with respect
to the DCD effect is left as free parameter. Our results can be
summarized as follows:

\begin{itemize}
\item{A 500 $\mic$ emissivity excess, with respect to the predictions
    of a modified black-body model with $\beta$=2, has been found in the
    peripheral parts of the
    Galactic plane ($35\degr<|l|<70\degr$) covered by the data. This
    excess can represent up to 16$\%$ to 20$\%$ of the total emission
    in some HII regions.}

\item{The dust temperature, derived from the TLS model, appears to
    be slightly warmer in the central (by 1.4 K, for $|l|<35\degr$) than the peripheral
    Galactic regions covered by the data. Regions near the Galactic center have
    temperatures in the range 17-25 K, whereas the median temperature
    across the rest of the Galactic plane covered is close to
    18.6 K.}

\item{We have found a flattening of the emissivity spectra in the range 100-500 $\mic$ 
  with increasing dust temperature. A model using a mixture of temperatures along the
  LOS has been used to verify wether this could be responsible for the
  observed behaviour. The results strongly suggest that the changes
  in the observed emissivity spectra with dust 
temperature cannot be accounted for by an LOS effect only, and that
they must be caused instead by intrinsic variations in the dust
properties with environment. These results are indeed compatible with
the predictions of the TLS model with dust temperature, and suggest variations in
   the degree of amorphisation of the grains, i. e.  the disorder at the
      atomic and/or nanometer scales.}

\item{The 500 $\mic$ emissivity excess can be explained by the
    intensity produced by the TLS processes. Indeed, the spatial
  variations in the $A$ parameter (i.e. the amplitude of the TLS effects) follow the distribution of the
  excess.} 
\begin{figure*}[!t]
\begin{center}
\includegraphics[width=18cm]{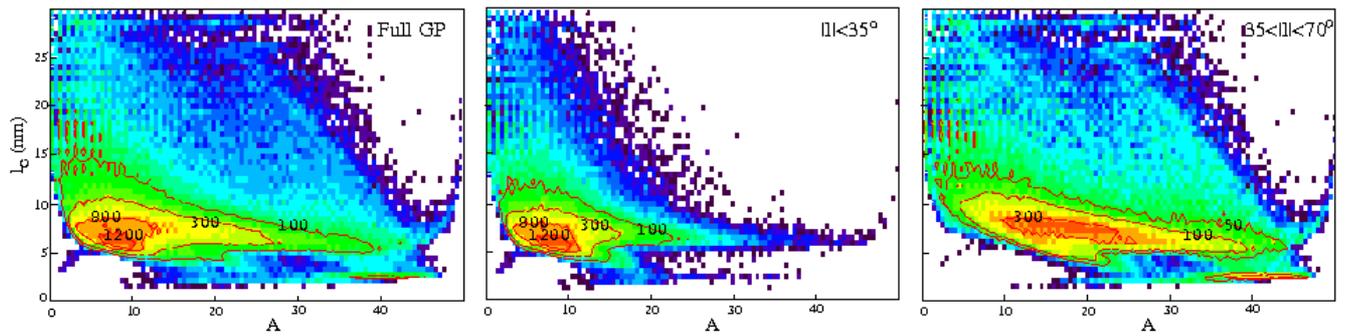}
\caption{Best-fit model parameters $l_c$ as a function of $A$, derived for
  the TLS model, across the entire Galactic plane (left panel), in the
  central (middle panel) and the peripheral regions of the Galactic plane (right panel). The overplotted
  contours represent the density of pixels, with levels at 100, 300,
  800, and 1200 for all the Galactic plane and the central regions, and levels at 50, 100,
  and 300 for the peripheral regions of the Galactic plane. \label{fig_A_lc}}
\end{center}
\end{figure*}

\item{Dust properties along the Galactic plane seem to differ
  from those of the solar neighborhood, the excess being smaller in
  the latter than expected from an extrapolation of the Galactic plane
  behavior.}

\item{The spatial distribution of $l_c$ (i.e. the correlation length of the DCD effect) does not present significant
    variations along the Galactic plane. Statistically, $l_c$ is three times
    lower than in the solar neighborhood, although it is close to the value
    obtained for compact sources.}

\item{Results in the framework of the TLS model indicate that dust
    grains are characterized by a degree of
    amorphization that is larger along the Galactic plane than in the solar neighborhood.
    In particular, the degree of
    amorphization tends to increase in the peripheral part of the
    plane covered by the Hi-GAL data. This work shows that specific
    variations in the mechanical structure of the material
    constituting the grains are likely to vary
    from the Galactic center towards the peripheral regions of the
    plane. }

\item{Variations in the dust temperature and the 500 $\mic$
    emissivity excess as a function of Galactic longitude appear to
  correlate with the locations of Galactic spiral arms. In the context
of the TLS model, this may suggest that changes in the mechanical
structure of the grains occur in the spiral arms.}

\end{itemize} 

The combination of Planck and Herschel data will provide the opportunity 
to improve the current constraints on the dust properties. 

\begin{acknowledgements}
We would like to thank the anonymous referee for his invaluable
comments. D. P. acknowledges grant support from the Centre National
d'Etudes Spatiales (CNES). 
\end{acknowledgements}

\end{document}